\documentclass{article}

\usepackage{microtype}
\usepackage{graphicx}
\usepackage{subcaption}
\usepackage{booktabs}
\usepackage{float}
\usepackage{amsmath}
\usepackage{multirow}
\usepackage{tikz}
\usepackage{pgfplots}
\pgfplotsset{compat=1.18} 

\usepackage[accepted]{icml2026}

\usepackage{hyperref}

\usepackage{algorithm}

\begin{document}

\twocolumn[
  \icmltitle{Automated Malware Family Classification using Weighted Hierarchical Ensembles of Large Language Models}



  \icmlsetsymbol{equal}{*}

  \begin{icmlauthorlist}
    \icmlauthor{Samita Bai}{yyy}
    \icmlauthor{Hamed Jelodar}{yyy}
    \icmlauthor{Tochukwu Emmanuel Nwankwo}{yyy}
    \icmlauthor{Parisa Hamedi}{yyy}
    \icmlauthor{Mohammad Meymani}{yyy}
    \icmlauthor{Roozbeh Razavi-Far}{yyy}
    \icmlauthor{Ali A. Ghorbani}{yyy}
\end{icmlauthorlist}

\icmlaffiliation{yyy}{Canadian Institute for Cybersecurity, Faculty of Computer Science, University of New Brunswick, Fredericton, Canada}

\icmlcorrespondingauthor{Samita Bai}{samita.bai@unb.ca}
\icmlcorrespondingauthor{Ali A. Ghorbani}{ghorbani@unb.ca}
  \icmlkeywords{Machine Learning, ICML}

  \vskip 0.3in
]



\printAffiliationsAndNotice{}  

\begin{abstract}
  Malware family classification remains a challenging task in automated malware analysis, particularly in real-world settings characterized by obfuscation, packing, and rapidly evolving threats. Existing machine learning and deep learning approaches typically depend on labeled datasets, handcrafted features, supervised training, or dynamic analysis, which limits their scalability and effectiveness in open-world scenarios.

This paper presents a zero-label malware family classification framework based on a weighted hierarchical ensemble of pretrained large language models (LLMs). Rather than relying on feature-level learning or model retraining, the proposed approach aggregates decision-level predictions from multiple LLMs with complementary reasoning strengths. Model outputs are weighted using empirically derived macro-F1 scores and organized hierarchically, first resolving coarse-grained malicious behavior before assigning fine-grained malware families. This structure enhances robustness, reduces individual model instability, and aligns with analyst-style reasoning.

The framework is evaluated on the SBAN (Source-Binary-Assembly-Natural Language) dataset, which includes real-world Windows PE malware samples spanning multiple families and obfuscation techniques. Experimental results demonstrate that the weighted hierarchical ensemble consistently outperforms individual LLMs and unweighted aggregation methods under constrained conditions where labeled data, dynamic execution, and fine-tuning are unavailable, highlighting its practicality for open-world malware analysis.
\end{abstract}

\section{Introduction}

The rapid growth and increasing sophistication of malware pose persistent challenges to cybersecurity. More than 450,000 new malware samples are registered daily, and the total number of known instances has increased nearly fivefold over the past decade~\cite{AVTest}. Widespread use of packing, obfuscation, and polymorphism limits the effectiveness of traditional signature-based and heuristic detection~\cite{Christ_2004,You_2011,Souri_2018}, making malware family classification a critical capability for automated analysis and threat intelligence~\cite{Nataraj_2011,Raff_2017,Abusitta_2021}.

Prior machine learning and deep learning approaches for malware family classification rely on handcrafted features, extensive labeled data, or costly dynamic analysis, and often degrade under obfuscation and concept drift~\cite{Shalag_2018,Ma_2021,Saxe_2015,Anderson_2018,Guldemir_2025}. These limitations hinder generalization to unknown or evolving malware families.

Recent large language models (LLMs) demonstrate strong capabilities in code understanding and zero-shot reasoning, showing promise for malware analysis tasks~\cite{Kojima_2022,Hui_2024,Espejel_2023,Jelodar_2025}. However, individual LLMs produce inconsistent predictions across malware families and prompt variations, and existing LLM-based pipelines typically rely on single-model inference or flat ensembling, limiting robustness and reliability~\cite{Ma_2021,Anderson_2018,Zhang_2025}.

To address these challenges, we propose a malware family classification framework based on \emph{weighted hierarchical ensembles of large language models}. The approach operates at the decision level by aggregating predictions from multiple pretrained LLMs without labeled training data, fine-tuning, or dynamic execution. Model contributions are weighted using macro-F1 performance, and predictions are organized within a behavior-aware hierarchical decision structure that reflects coarse-grained malware behaviors followed by fine-grained family attribution.

Experiments on a real-world PE malware dataset show that weighted hierarchical ensembling consistently outperforms individual LLMs and unweighted aggregation strategies under constrained settings. By decomposing classification into hierarchical decision stages, the proposed framework improves robustness, mitigates semantic ambiguity between families, and aligns with analyst-driven malware reasoning.

\textbf{Research Contributions}
The main contributions of this paper are summarized as follows:

\begin{enumerate}
    \item \textbf{Decision-Level Malware Family Classification using LLMs}  
    Malware family classification is formulated as a decision aggregation problem over pretrained large language models, eliminating dependence on handcrafted feature pipelines, supervised retraining, and dynamic execution.

    \item \textbf{Performance-Aware Weighted Ensemble Strategy}  
    A macro-F1–based weighting scheme is introduced to combine complementary LLM predictions, improving robustness across malware families and reducing the impact of individual model instability.

    \item \textbf{Hierarchical Decision Framework for Family Attribution}  
    A hierarchical ensemble structure is designed to first distinguish coarse-grained malware behaviors and subsequently resolve fine-grained family labels, reducing ambiguity in overlapping family boundaries.

    \item \textbf{Comprehensive Evaluation under Realistic Constraints}  
    Extensive experiments compare individual LLMs, unweighted ensembles, and the proposed weighted hierarchical ensemble, demonstrating consistent performance gains without reliance on labeled training data or dynamic analysis.
\end{enumerate}

\textbf{Organization}
The remainder of this paper is organized as follows. Section~2 reviews prior work on malware family classification and LLM-based malware analysis. Section~3 describes the proposed weighted hierarchical LLM ensemble framework. Section~4 presents the experimental setup and evaluation results. Section~5 concludes the paper and discusses future research directions.

\section{Background and Related Work}

Malware family classification is a fundamental and challenging problem in malware analysis, particularly for Windows Portable Executable (PE) files. Unlike binary detection, which separates benign from malicious samples, family classification aims to attribute malware to semantically meaningful families that exhibit overlapping behaviors, evolving implementations, and ambiguous boundaries.

Early signature-based and heuristic methods scale poorly and fail under modern malware characteristics such as packing, obfuscation, and polymorphism~\cite{Abusitta_2021}. Subsequent work introduced static and dynamic feature extraction, representation learning, and graph-based reasoning, but these approaches rely on handcrafted pipelines, labeled data, or costly dynamic execution. A detailed survey of classical malware family classification methods is provided in Appendix~\ref{app:relatedWork}.

Recent work has explored higher-level semantic abstractions using Large Language Models (LLMs), motivating their application to malware analysis and attribution.

\subsection{Large Language Models for Malware Analysis}

LLMs have emerged as powerful tools for program analysis due to their ability to capture semantic relationships beyond surface-level syntax. Pretrained on large code and natural-language corpora, LLMs can infer program intent, reason about API usage, and summarize functionality, making them attractive for malware analysis tasks.

Prior studies show that LLMs exhibit strong semantic understanding but unstable performance under obfuscation and code transformations~\cite{Fang, jelodar2025large}. Benchmarking work demonstrates that LLMs can infer malicious semantics but suffer from inconsistent predictions across malware families~\cite{He, llmMalwareSurvey2025}. LLMs have also been applied to malware comprehension, deobfuscation, and behavior inference~\cite{llmDeobfuscation2024, llmRE2024, binaryAttribution2024}.

Despite these advances, existing LLM-based approaches remain limited by prediction instability, sensitivity to prompt design and obfuscation, and reliance on fine-tuning or domain supervision. These limitations indicate that single-model LLM pipelines are insufficient for robust malware family classification in open-world settings.

\subsection{Weighted Hierarchical Ensembles of Large Language Models}

Individual LLMs exhibit complementary strengths and weaknesses, motivating ensemble learning to improve robustness. Prior work shows that ensembling multiple LLMs improves reliability over single-model inference, even without fine-tuning~\cite{sunLLMEnsemble2025, majorityRules2025}. Performance-aware weighting schemes further enhance ensemble stability by assigning higher influence to more reliable models based on validation performance~\cite{llmEnsembleSurvey2025}.

Malware families share higher-level behavioral characteristics, and hierarchical decision structures have been shown to improve robustness and interpretability under ambiguous class boundaries~\cite{Abusitta_2021, UcciSurvey}. Hierarchical classification decomposes attribution into coarse-grained behavioral reasoning followed by fine-grained family resolution.

In contrast to prior work, our approach introduces a weighted hierarchical ensemble of LLMs for malware family classification from source code. Predictions from multiple pretrained LLMs are aggregated at the decision level using empirically derived macro-F1 weights. A behavior-aware hierarchical structure enables coarse-to-fine reasoning, mitigating instability and semantic ambiguity.

Unlike existing LLM-based malware analysis methods that rely on single models or flat aggregation, the proposed framework demonstrates that principled, domain-aware ensembling yields improved robustness and accuracy under constrained, zero-label settings.

\section{Methodology}
The proposed framework treats malware family classification as a structured decision-making problem rather than a flat label prediction task, enabling explicit modeling of semantic overlap and behavioral ambiguity between malware families.

This section describes our methodology for malware family classification using a reliability-aware ensemble of large language models (LLMs). The proposed framework aggregates independent LLM predictions over malware source code using a gold-calibrated weighted hierarchical decision process that integrates behavioral consistency and domain-specific specificity.

\begin{figure*}[t]
    \centering
    \includegraphics[width=\textwidth]{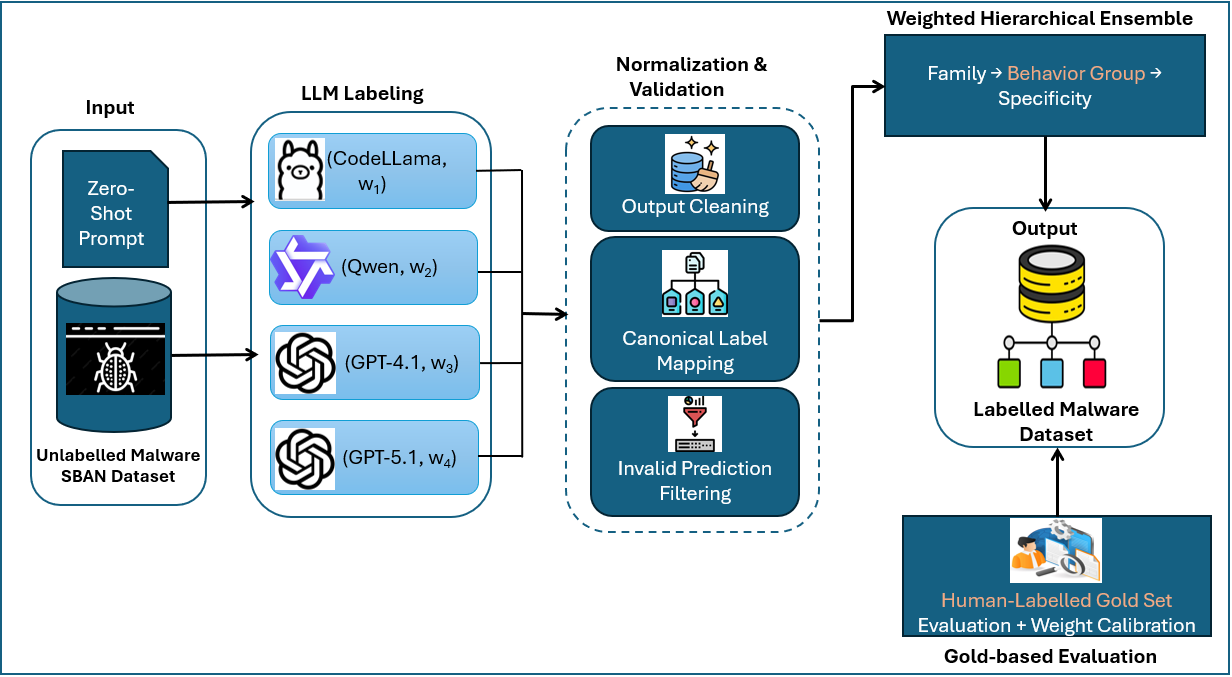}
    \caption{Proposed zero-shot LLM ensemble pipeline for malware family classification. A shared classification prompt is applied to multiple LLMs, followed by normalization, weighted hierarchical ensembling, and gold-based evaluation.}
    \label{fig:pipeline}
\end{figure*}

Figure~\ref{fig:pipeline} illustrates the full pipeline of the proposed malware family classification framework. Given an unlabelled malware source code dataset, each sample is paired with a fixed zero-shot classification prompt and independently processed by multiple LLMs. All models operate under identical prompting conditions to ensure fair and directly comparable outputs.

The raw predictions produced by the LLMs are subsequently passed through a normalization and validation stage, which removes invalid responses, maps synonymous labels to canonical malware families, and enforces consistency with the predefined taxonomy. The validated predictions are then combined using a weighted hierarchical ensemble that aggregates model outputs at the family level, refines decisions using behavior-group consensus, and resolves remaining ambiguities via specificity-based tie-breaking.

The final output of the pipeline is a single malware family label per sample. A human-labeled gold set is incorporated only during evaluation and for calibrating model weights and does not influence the inference process.

\begin{figure}[ht]
    \vskip 0.1in
    \begin{center}
    \centerline{\includegraphics[width=\columnwidth]{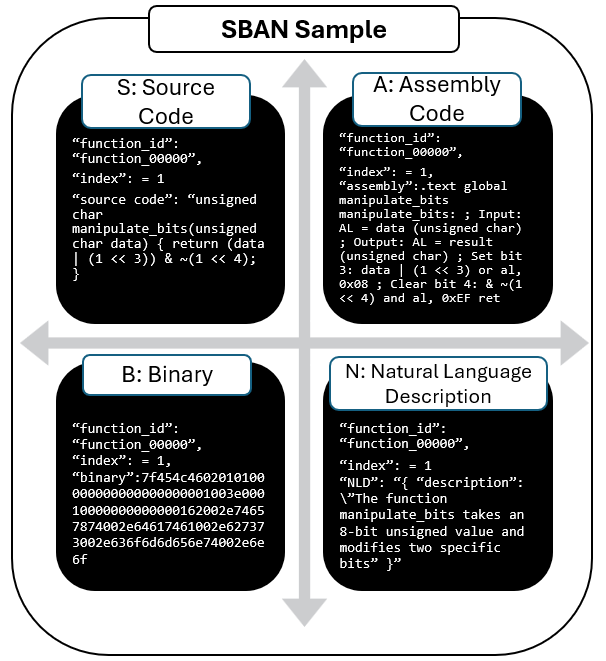}}
    \caption{A sample from SBAN dataset.}
    \label{fig:wide_figure}
    \end{center}
\end{figure}

\subsection{Problem Formulation}
\label{sec:problem_formulation}

We formulate malware family classification as a zero-label, decision-level aggregation problem over pretrained large language models (LLMs). The goal is to assign each malware sample to a single canonical family without relying on supervised training, feature-level learning, dynamic execution, or model fine-tuning.

Let $\mathcal{X}$ denote the input space of malware representations, where each sample $x \in \mathcal{X}$ consists of static source-code content and associated API import information extracted from a Windows Portable Executable (PE) program. Let $\mathcal{Y} = \{f_1, f_2, \ldots, f_{10}\}$ denote the fixed set of ten canonical malware families defined in Table~1.

We are given a collection of $N$ pretrained large language models $\{M_1, M_2, \ldots, M_N\}$. Each model $M_i$ maps an input sample $x$ to a predicted malware family label:
\begin{equation}
    p_i(x) = M_i(x), \quad p_i(x) \in \mathcal{Y}.
\end{equation}
All models operate under a shared zero-shot classification prompt and identical input preprocessing to ensure that observed performance differences arise solely from model behavior rather than prompt engineering or representation effects.

The task is to define an ensemble decision function $\mathcal{E}$ that aggregates the individual model predictions into a single final family label:
\begin{equation}
    \hat{y}(x) = \mathcal{E}\big(\{p_i(x)\}_{i=1}^{N}, \{w_i\}_{i=1}^{N}\big),
\end{equation}
where $w_i \geq 0$ denotes the reliability weight assigned to model $M_i$ and $\sum_{i=1}^{N} w_i = 1$.

In contrast to conventional supervised classification, no labeled training data are used to fit model parameters or to learn feature representations. Instead, a small human-labeled gold set $G = \{(x_j, y_j)\}_{j=1}^{m}$ is employed exclusively for (i) estimating model reliabilities and calibrating the weights $\{w_i\}$ and (ii) evaluating classification performance. The gold set does not influence inference-time predictions for unseen samples.

The framework operates under zero-label inference, static-only analysis, a fixed malware taxonomy, and a single invariant zero-shot prompt.
Given these constraints, the core challenge is to design an ensemble decision function $\mathcal{E}$ that is robust to (i) inter-model disagreement, (ii) semantic overlap between malware families, and (iii) instability arising from obfuscation, truncation, and prompt sensitivity. To address this, we introduce a weighted hierarchical aggregation strategy that decomposes attribution into coarse-grained behavioral reasoning followed by fine-grained family resolution, as described in the following sections.

\subsection{Dataset Used: SBAN}

This study uses the \textbf{SBAN (Source-Binary-Assembly-Natural language)} dataset~\cite{SBAN_Jelodar_2025}, a large-scale multimodal corpus aligning source code, assembly, binaries, and natural-language descriptions. SBAN contains over 3.3 million samples, including approximately 2.95~M benign programs from \textit{CodeNet} and \textit{ExeBench}, and about 0.67~M malware samples from \textit{BODMAS}, \textit{MalwareBazaar}, \textit{SOREL-20M}, \textit{DIKE}, and \textit{xLangKode}.

We use the malware subset for family-level clustering and classification in a zero-label setting. Since SBAN does not provide malware family labels, it is well-suited for unsupervised learning based on code semantics. Additional datasets are described in Appendix~\ref{app:datasets}.

\begin{table}[!t]
\centering
\caption{Malware families used for classification}
\label{tab:families}
\begin{tabular}{p{0.4\columnwidth} p{0.5\columnwidth}}
\hline
\textbf{Family} & \textbf{Description} \\ \hline
Trojan & Generic malware performing malicious actions without self-propagation \\
Worm & Self-replicating malware that spreads autonomously \\
Virus & Malware attaching to host executables for propagation \\
Ransomware & Malware that encrypts or locks data for extortion \\
Backdoor / Remote Access Trojan & Malware enabling persistent remote control \\
Dropper & Malware whose primary role is payload installation \\
Downloader & Malware that retrieves and executes external payloads \\
Packed / Obfuscated Malware & Malware characterized by concealment techniques \\
Spyware / Infostealer & Malware that covertly collects sensitive information \\
Bot / Botnet Client & Malware participating in coordinated botnet activity \\
\hline
\end{tabular}
\end{table}

\subsection{Malware Family Taxonomy}
We formulate malware family classification as a multi-class problem over a fixed set of ten canonical malware families. Each malware sample is assigned exactly one family label. Table~\ref{tab:families} summarizes the taxonomy used throughout this study. This taxonomy is fixed across all experiments and serves as the canonical label space for normalization, ensemble inference, and evaluation.

\subsection{Zero-Shot Prompt Design}
\label{sec:prompt}
\begin{figure}[H]
  \centering
  \includegraphics[width=\columnwidth]{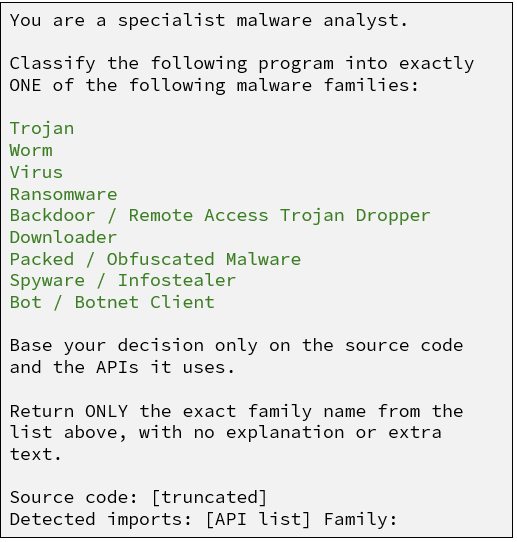}
  \caption{Zero-shot malware family classification prompt used to elicit decision-level predictions from pretrained large language models.}
  \label{fig:prompt}
\end{figure}

We employ a zero-shot prompting strategy for all experiments. A single, fixed classification prompt as shown in Figure~\ref{fig:prompt} is used uniformly across all LLMs, and no in-context examples or model-specific prompt variants are applied. This ensures that any observed performance differences arise solely from model behavior rather than prompt engineering effects. The prompt instructs the model to classify a malware program into exactly one of ten canonical malware families. It explicitly restricts the output format to the family name only and prohibits explanations or additional text. The same prompt is used across all LLMs, and the input source code is truncated to a fixed maximum length where necessary to meet model input constraints. The prompt explicitly restricts outputs to the predefined taxonomy, reducing hallucinated or non-canonical responses.

\subsection{Large Language Models}
We initially evaluated multiple LLMs for malware family labeling, including Qwen, CodeLLaMA, DeepSeekCoder, GPT-4.1, and GPT-5.1. Based on preliminary evaluation using a human-labeled gold set, DeepSeekCoder-coder-6.7b-instruct was excluded due to consistently poor accuracy (less than 10\%) and macro-F1 performance.
The final ensemble consists of the following four models:
\begin{itemize}
    \item Qwen3-4B-Instruct-2507
    \item CodeLLaMA-7B-Instruct-hf
    \item GPT-4.1
    \item GPT-5.1
\end{itemize}

Each model independently predicts a malware family label for every sample.

\subsection{Normalization and Validation}
Raw LLM outputs frequently exhibit inconsistencies such as synonymous labels
(e.g., ``Backdoor'', ``RAT''), formatting artifacts, or invalid responses. We apply a normalization and validation layer that:
\begin{enumerate}
    \item Maps synonymous and variant outputs to canonical family names;
    \item Removes invalid predictions (empty responses, explanations, markdown);
    \item Treats \emph{Trojan} and \emph{Backdoor / Remote Access Trojan} as
    equivalent \emph{during evaluation} due to their close semantic overlap.
\end{enumerate}
Only validated predictions mapped to the canonical taxonomy are forwarded to the ensemble.

\subsection{Hierarchical Behavior-Aware Ensemble}
\label{subsec:hierarchical_ensemble}

Large Language Models often disagree on malware family predictions, particularly for semantically overlapping categories such as \textit{Trojan}, \textit{Backdoor}, and \textit{Spyware}. To derive a single reliable label from multiple heterogeneous LLM outputs, we propose a weighted hierarchical ensemble that integrates majority voting by reliability, behavior-group consolidation, and specificity-based arbitration.

\subsubsection{Stage 1: Weighted Family Voting}
The complete gold-calibrated weighted hierarchical ensembling procedure is summarized in Algorithm~\ref{alg:ensemble}.
Let \(P = \{p_1, \ldots, p_N\}\) denote the normalized predictions produced by
\(N\) LLMs, and let \(w_i\) be the reliability weight of model \(i\).
For each family \(f\), a weighted score is computed as:

\begin{equation}
S(f) = \sum_{i : p_i = f} w_i.
\label{eq:score}
\end{equation}

If any family satisfies:
\begin{equation}    
S(f) > \frac{1}{2} \sum_{i=1}^{N} w_i,
\end{equation}

The ensemble outputs that label directly.
\\
\subsubsection{Stage 2: Behavior-Group Consolidation}

If no family achieves a weighted majority, predictions are grouped into coarse behavioral categories based on operational similarity, as shown in
Table~\ref{tab:behavior_groups}.

Weighted scores are aggregated at the group level. If a behavior group exceeds the weighted majority threshold, the highest-scoring family within that group is selected.

\subsubsection{Stage 3: Specificity-Based Tie Resolution}

When ties persist within a dominant group, or no group achieves a majority, we apply a predefined specificity ranking that favors behaviorally informative
families (e.g., \textit{Backdoor} over \textit{Trojan},
\textit{Downloader} over \textit{Dropper}). 

The final family is selected as:
\begin{equation}
f^{*} = \arg\min_{f \in T} R(f),
\end{equation}
where \(T\) denotes the set of tied candidates and \(R(f)\) the specificity rank.

\subsection{Hierarchy Design Rationale}
\label{sec:hierarchy_rationale}

The hierarchical decision structure in Section~3.6 addresses systematic inter-model confusion among semantically adjacent malware families, particularly within the Trojan--Backdoor--Spyware cluster and the Dropper--Downloader pair. These ambiguities arise from overlapping behavioral semantics under static-only analysis, making flat aggregation unstable.

We partition the ten canonical families into five coarse-grained behavior groups (Table~2) based on dominant operational characteristics, reflecting analyst-driven attribution practice and emphasizing behavioral intent.

The ensemble follows a coarse-to-fine process: when no family achieves a strict weighted majority, predictions are first aggregated at the behavior-group level to suppress spurious minority labels and stabilize decisions.

At both the family and group stages, we use a strict weighted-majority threshold of $tfrac{1}{2} \sum_i w_i$. Ties within a dominant group are resolved using a fixed specificity ranking that favors behaviorally informative labels.

Overall, the hierarchy improves robustness and interpretability by reducing inter-model instability under constrained, zero-label conditions.

\begin{table}[!t]
  \caption{Mapping of malware families to behavior groups.}
  \label{tab:behavior_groups}
  \begin{center}
    \begin{small}
      \begin{sc}
        \begin{tabular}{p{0.50\columnwidth} p{0.30\columnwidth}}
          \toprule
          \textbf{Families} & \textbf{Behavior Group} \\
          \midrule
          Trojan, Backdoor/RAT, Spyware/Infostealer, Bot & TROJAN\_LIKE \\
          Dropper, Downloader & INSTALLER \\
          Worm, Virus & SELF\_REPLICATING \\
          Ransomware & RANSOMWARE \\
          Packed / Obfuscated Malware & OBFUSCATED \\
          \bottomrule
        \end{tabular}
      \end{sc}
    \end{small}
  \end{center}
  \vskip -0.1in
\end{table}

\subsection{Gold Set Construction and Weight Calibration}
To calibrate model reliabilities, we construct a human-labeled gold set. The gold-standard dataset consists of 200 malware samples manually labeled by multiple malware analysis experts following a predefined family taxonomy. Inter-annotator agreement was measured using Cohen’s $\kappa$, indicating substantial agreement. Disagreements were resolved through consensus discussion. Full labeling protocol and statistics are provided in Appendix~\ref{app:goldset}.

For each model \(M_i\), we compute its classification accuracy on the gold set, counting Trojan and Backdoor/RAT predictions as equivalent. Model weights are derived via linear normalization:
\begin{equation}
w_i = \frac{\text{Acc}(M_i)}{\sum_j \text{Acc}(M_j)}.
\end{equation}

The gold set is used exclusively for weight calibration and evaluation and is never used during ensemble inference on unlabeled samples.

\begin{algorithm}[t]
\caption{Gold-Calibrated Weighted Hierarchical Ensemble}
\label{alg:ensemble}
\small
\textbf{Input:} Malware source code sample $x$; predictions $\{y_i\}$ from models $\{M_i\}$; weights $\{w_i\}$.\\
\textbf{Output:} Final malware family label $y^\ast$.
\vspace{2pt}

\begin{enumerate}\setlength\itemsep{1pt}
\item Normalize and validate all predictions $y_i$.
\item Remove invalid or unmappable outputs.
\item Compute weighted family scores: for each family $f$, set $S(f) \gets \sum_{i: y_i=f} w_i$.
\item If $\max_f S(f) > \frac{1}{2}\sum_i w_i$, return $\arg\max_f S(f)$.
\item Group predictions by behavior class and compute weighted behavior-group scores.
\item If $\max_g S(g) > \frac{1}{2}\sum_i w_i$:
  \begin{enumerate}\setlength\itemsep{1pt}
  \item Select the highest-scoring family within the winning group.
  \item If tied, resolve using specificity ranking.
  \item Return the selected family.
  \end{enumerate}
\item Resolve remaining ties using global specificity ranking and return the final family label.
\end{enumerate}
\end{algorithm}

\section{Experimental Setup and Evaluation Results}

This section presents the experimental setup and evaluation of the proposed automated malware family classification framework based on weighted and hierarchical ensembles of large language models (LLMs). The evaluation assesses whether ensembling improves robustness and classification performance relative to individual LLM baselines, and quantifies the contribution of weighting and hierarchical decision structures.

\subsection{System Configuration}

All experiments were conducted on a high-performance computing server equipped with an NVIDIA H100 NVL GPU. The system is powered by dual Intel Xeon Gold 6548Y+ processors (2 sockets, 32 cores per socket, 64 threads per socket), providing a total of 128 hardware threads. The CPUs operate under the x86\_64 architecture with 32/64-bit modes and support 46-bit physical and 57-bit virtual addressing. The system uses a little-endian memory model.

\subsection{Dataset and Evaluation Protocol}

Experiments were conducted on a manually curated gold-standard dataset consisting of 200 malware samples with verified family labels spanning ten canonical malware families. Each sample was independently classified by multiple LLMs, and all predictions were normalized to the canonical label set before evaluation.

Performance was measured using Accuracy, Macro Precision, Macro Recall, and Macro F1-score. Macro-averaged metrics were emphasized due to class imbalance and to ensure equal importance across malware families.

Due to the limited size of the gold-standard dataset, results are reported deterministically without statistical significance testing.
The focus is on relative performance trends across models and ensemble variants.

\subsection{Individual LLM Baselines}

Four LLMs were evaluated independently as single-model baselines: Qwen, CodeLLaMA, GPT-4.1, and GPT-5.1. Table~\ref{tab:individual_results} reports their performance.

\begin{table}[h]
\centering
\caption{Performance of Individual LLMs on Malware Family Classification}
\label{tab:individual_results}
\begin{tabular}{p{0.22\columnwidth} p{0.13\columnwidth} p{0.13\columnwidth} p{0.13\columnwidth} p{0.10\columnwidth}}
\hline
\textbf{Model} & \textbf{Accuracy} & \textbf{Macro-P} & \textbf{Macro-R} & \textbf{Macro-F1} \\
\hline
Qwen        & 0.695 & 0.492 & 0.325 & 0.380 \\
CodeLLaMA  & 0.420 & 0.278 & 0.238 & 0.159 \\
GPT-4.1    & 0.710 & 0.416 & 0.422 & 0.403 \\
GPT-5.1    & 0.675 & 0.476 & 0.437 & 0.427 \\
\hline
\end{tabular}
\end{table}

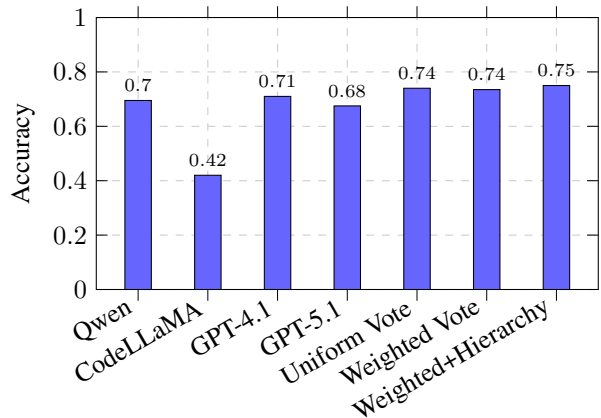
\begin{figure}[t]
\centering
\begin{tikzpicture}
\begin{axis}[
    ybar,
    bar width=10pt,
    width=\linewidth,
    height=5.2cm,
    ymin=0,
    ymax=1.0,
    ylabel={Accuracy},
    symbolic x coords={
        Qwen,
        CodeLLaMA,
        GPT-4.1,
        GPT-5.1,
        Uniform Vote,
        Weighted Vote,
        Weighted+Hierarchy
    },
    xtick=data,
    x tick label style={rotate=30, anchor=east},
    nodes near coords,
    nodes near coords align={vertical},
    every node near coord/.append style={font=\scriptsize},
    grid=both,
    major grid style={dashed,gray!40},
    axis line style={black},
    tick style={black},
    title={Accuracy Comparison of Individual Models and Ensembles}
]
\addplot[
    fill=blue!60,
    draw=black
] coordinates {
    (Qwen,0.695)
    (CodeLLaMA,0.420)
    (GPT-4.1,0.710)
    (GPT-5.1,0.675)
    (Uniform Vote,0.740)
    (Weighted Vote,0.735)
    (Weighted+Hierarchy,0.750)
};
\end{axis}
\end{tikzpicture}
\caption{Accuracy comparison of individual LLMs and ensemble strategies on the 200-sample gold standard. The final weighted hierarchical ensemble achieves the highest overall accuracy.}
\label{fig:accuracy_comparison}
\end{figure}

Among individual models, GPT-4.1 achieved the highest accuracy, while GPT-5.1 provided the strongest Macro-F1 score, indicating better balance across classes. CodeLLaMA consistently underperformed and was therefore excluded from subsequent ensemble configurations.
As shown in Fig.~\ref{fig:accuracy_comparison}, ensemble strategies consistently outperform individual LLMs in terms of accuracy.

\subsection{Ensemble Configurations}

Three ensemble configurations were evaluated:
\begin{itemize}
        \item \textbf{Uniform Majority Voting (Row 1):} Each model contributed equally, and the final family label was selected by simple majority vote.
    \item \textbf{Weighted Voting without Hierarchy (Row 2):} Model predictions were weighted based on validation-set Macro-F1 scores were computed on the gold standard, reflecting each model’s reliability across malware families. Weights were normalized to sum to one to prevent dominance by any single model
    \item \textbf{Weighted Hierarchical Ensemble (Row 3):} A two-stage decision process was employed, where coarse-grained family groups were resolved first, followed by fine-grained family selection using confidence-weighted aggregation.
\end{itemize}

\subsection{Result Analysis}

Individual LLMs show high performance variability, and although GPT-4.1 and GPT-5.1 perform best among single models, none achieves consistently strong Macro-F1 across families. Uniform majority voting improves over the strongest single model, confirming complementary model behavior, while weighted voting without hierarchy slightly underperforms uniform voting.
The weighted hierarchical ensemble achieves the best performance, improving accuracy and recall while maintaining strong Macro-F1, and enhances interpretability through intermediate coarse-grained decisions.

\subsection{Ablation Study}
An ablation analysis confirms the contribution of each ensemble component. Uniform voting establishes a strong baseline, weighted voting improves precision through model calibration, and hierarchical decision-making further enhances robustness by resolving coarse-grained family ambiguity before fine-grained classification. The combination of weighting and hierarchy yields the most balanced and stable performance across all metrics.
The initial hierarchical configuration reflects a naive hierarchy without confidence gating, while the final variant incorporates refined confidence thresholds and suppression of low-confidence branches.
Table~\ref{tab:ablation} summarizes the performance of the ensemble variants.

\begin{table}[!t]
  \caption{Ablation Study of Ensemble Strategies}
  \label{tab:ablation}
  \begin{center}
    \begin{small}
      \begin{tabular}{p{0.30\columnwidth} p{0.10\columnwidth} p{0.10\columnwidth} p{0.10\columnwidth} p{0.10\columnwidth}}
        \toprule
        \textbf{Method} & \textbf{Accuracy} & \textbf{Macro-P} & \textbf{Macro-R} & \textbf{Macro-F1} \\
        \midrule
        Uniform Voting (Row 1) & 0.740 & 0.546 & 0.464 & 0.481 \\
        Weighted Voting (Row 2) & 0.735 & 0.527 & 0.464 & 0.469 \\
        Weighted + Hierarchy (Initial) & 0.725 & 0.507 & 0.462 & 0.460 \\
        \textbf{Weighted + Hierarchy (Final)} & \textbf{0.750} & \textbf{0.530} & \textbf{0.479} & \textbf{0.481} \\
        \bottomrule
      \end{tabular}
    \end{small}
  \end{center}
  \vskip -0.1in
\end{table}

\subsection{Discussion}

Uniform majority voting (Row~1) outperformed individual LLM baselines, confirming that ensemble diversity captures complementary semantic signals. The weighted voting scheme without hierarchy (Row~2) achieved similar performance, indicating limited gains from weighting alone.

The initial hierarchical ensemble underperformed due to error propagation at uncertain coarse stages. After refinement with confidence-based weighting and selective routing, the final weighted hierarchical ensemble achieved the best accuracy and competitive Macro-F1.

These results show that hierarchical ensembling is effective only when combined with calibrated weighting. Despite dataset limitations, the consistent gains demonstrate the robustness of the proposed design.

\section{Conclusion}

This paper presented a decision-level malware family classification framework based on weighted hierarchical ensembles of large language models. The method aggregates predictions from multiple pretrained LLMs without labeled training data, fine-tuning, or dynamic analysis.

Experiments on a gold-standard subset of the SBAN dataset show that ensemble strategies consistently outperform individual LLMs. While uniform voting yields notable gains, the refined weighted hierarchical ensemble achieves the strongest performance by combining calibrated model weighting with a structured decision process, improving robustness, class balance, and interpretability under zero-label settings. The framework is evaluated on a relatively small dataset and relies on pretrained LLM outputs and a manually defined hierarchy, which may limit generalization and introduce routing errors. In addition, the current design focuses only on static code inputs.

Future work will scale evaluation to larger corpora, explore data-driven hierarchy induction, and incorporate lightweight dynamic or hybrid signals. Overall, this work demonstrates that decision-level LLM ensembling offers a practical and interpretable approach to scalable malware family classification.

\nocite{langley00}

\bibliography{example_paper}
\bibliographystyle{icml2026}

\newpage
\appendix
\onecolumn

\section{Classical and Representation-based Malware Family Classification}
\label{app:relatedWork}
This appendix provides a detailed survey of classical and representation-based malware family classification techniques that form the historical and methodological foundation of automated malware analysis. These approaches rely on static and dynamic feature extraction, handcrafted representations, deep learning, and graph-based reasoning. While they have achieved strong performance under controlled conditions, they exhibit fundamental limitations in open-world environments characterized by obfuscation, concept drift, and limited labeled data.

\subsection{AV Labeling Approaches}

Antivirus (AV) labeling has been widely adopted to derive malware family labels by aggregating outputs from multiple security vendors. Tools such as AVClass~\cite{ref2} and AVClass2~\cite{AVClass2} normalize noisy AV labels using heuristic mappings and co-occurrence statistics, enabling large-scale family annotation. More recently, Joyce et al.~\cite{ref1} proposed ClarAVy, which applies a Variational Bayesian aggregation approach (SparseIBCC) to combine AV scan results more robustly.  
Although AV-based labeling is valuable for dataset construction and evaluation, it inherits vendor bias, label inconsistency, and limited coverage of emerging families, and therefore cannot serve as a reliable standalone family classification solution.

\subsection{Static Feature-Based Classification}

Static malware analysis extracts features from binaries without execution, such as PE header metadata, imported APIs, opcode sequences, byte n-grams, entropy measures, and section statistics~\cite{staticfeature}. These features are computationally efficient and have been widely used for malware family classification. For example, Ahmadi et al.~\cite{Ahmadi} employ byte-level and structural PE features to achieve high classification accuracy on the Microsoft Malware Challenge dataset, while Sun et al.~\cite{Sun} combine bytecode, assembly, and PE structural attributes using machine learning models.

Despite their effectiveness, static approaches are highly sensitive to packing, encryption, and obfuscation, which can drastically alter surface-level features while preserving malicious behavior~\cite{Abusitta_2021, staticPacking}. As a result, static feature-based classifiers often degrade significantly when deployed against real-world malware exhibiting polymorphic or metamorphic behavior.

\subsection{Dynamic and Hybrid Feature-Based Classification}

Dynamic analysis executes malware samples within sandboxed environments to observe runtime behavior, extracting features such as system-call traces, API invocation sequences, registry modifications, file system activity, and network communication~\cite{Abusitta_2021, Bensaoud}. These behavioral features capture semantic intent more robustly than static representations and are less affected by packing. For instance, Rieck et al.~\cite{Rieck} classify malware families based on behavioral system-call profiles, while Kolosnjaji et al.~\cite{Kolosnjaji} apply deep learning models to system-call sequences collected during execution.

Hybrid approaches combine static and dynamic features to exploit their complementary strengths. Studies such as~\cite{Firdausi, Islam} demonstrate improved classification accuracy by integrating structural information with runtime behavior. However, dynamic and hybrid methods incur high computational overhead, require specialized execution infrastructure, and are vulnerable to anti-sandbox and anti-VM evasion techniques, which limit scalability for large malware corpora.

\subsection{Advanced Representation Learning}

To mitigate the limitations of manual feature engineering, recent research has adopted representation-learning paradigms that learn discriminative malware characteristics directly from raw or structured inputs~\cite{Ma_2021, Abusitta_2021}. Image-based approaches transform binary content into visual representations suitable for convolutional neural networks, allowing models to capture structural patterns shared across malware families. Early work by Nataraj et al.~\cite{Nataraj_2011} introduced grayscale malware visualization, while subsequent studies~\cite{Saxe_2015, Eroglu2025, Zhao2023} demonstrated improved performance using deep CNNs and richer visual encodings. Despite their efficiency, image-based methods offer limited interpretability and degrade under heavy packing and obfuscation.

Binary-based deep learning approaches operate directly on raw executable bytes or PE structural fields without requiring disassembly. MalConv~\cite{Raff_2017} demonstrated end-to-end learning from raw binaries, while later work incorporated richer binary semantics and section-level context to improve robustness~\cite{Hao, Miraoui}. Nonetheless, these models require substantial labeled datasets and remain sensitive to adversarial code transformations.

Graph-based representations model malware structure using control-flow graphs (CFGs), call graphs, or API-dependency graphs~\cite{Kinable}. Recent work increasingly employs graph neural networks (GNNs) to learn expressive graph embeddings for malware family classification~\cite{Malhotra, Shokouhinejad}. Although graph-based reasoning captures high-level semantic behavior, it depends on reliable graph extraction, incurs high computational cost, and remains sensitive to obfuscation and incomplete disassembly, limiting applicability in open-world scenarios.

\subsection{Limitations of Classical and Representation-Based Methods}

Despite significant advances, classical, deep learning, and graph-based malware family classification techniques share common limitations. Most approaches depend heavily on labeled training data, are sensitive to obfuscation and family drift, incur substantial computational cost, and offer limited interpretability in real-world deployments. These challenges are particularly pronounced in open-world environments, where malware families evolve continuously and ground-truth labels are scarce.

These limitations motivate alternative approaches that shift the focus from feature-level learning to higher-level semantic decision-making mechanisms, enabling robust malware family classification without reliance on brittle feature extraction pipelines or expensive retraining.

\section{Datasets and Benchmarks}
\label{app:datasets}
A wide range of datasets have been developed for malware classification, each supporting different research objectives and levels of semantic detail (Table~\ref{tab:malware_datasets}). Traditional benchmarks such as EMBER \cite{Anderson} and SOREL-20M \cite{Harang} provide large-scale collections of PE files with static metadata and binary labels, enabling research on static detection but lacking curated family annotations or multimodal semantic context. BODMAS \cite{Yang} introduces well-defined family labels and temporal sampling, supporting studies on family drift, while MalNet \cite{Freitas} expands into graph-structured behavioral representations with millions of API-call graph samples for GNN-based analysis. Community repositories such as MalwareBazaar offer access to large numbers of real malware samples for collection and enrichment workflows, though labeling is inconsistent and not designed for supervised classification. Behavioral datasets such as Maldict \cite{Joyce} and updated multi-format datasets such as EMBER2024 provide more recent samples and challenge sets but remain metadata or feature-oriented and do not support high-level semantic reasoning.
In contrast to these feature- or structure-focused datasets, SBAN \cite{SBAN_Jelodar_2025} introduces a large-scale multimodal dataset aligned across binary, assembly, source-code, and natural-language levels, enabling semantic understanding, behavior-aware family attribution, and explainability. SBAN fills a key gap by supporting semantic malware analysis tasks such as variant clustering, retrieval-based reasoning, and LLM-assisted interpretation—capabilities that existing static or graph-only datasets cannot fully address.
\begin{table}[t]
  \caption{Overview of benchmark datasets used in malware classification research.}
  \label{tab:malware_datasets}
  \begin{center}
    \begin{small}
      \begin{sc}
        \renewcommand{\arraystretch}{1.15}
        \setlength{\tabcolsep}{2pt}
        \begin{tabular}{l c c c p{6.0cm}}
          \toprule
          Dataset & Size & Labels & Modalities & Strengths / Limitations \\
          \midrule

          EMBER (2018) \cite{Anderson} 
          & 1.1M 
          & Binary 
          & PE metadata 
          & Widely used benchmark for static ML detection; lacks family labels and semantic context. \\

          SOREL-20M (2020) \cite{Harang} 
          & 20M 
          & Weak tags 
          & Metadata 
          & Large-scale dataset enabling detection research; lacks curated family structure or multimodal signals. \\

          BODMAS (2021) \cite{Yang} 
          & 134K 
          & 581 fam. 
          & Struct. feats. 
          & Supports temporal drift and family-level evaluation; limited modality diversity and no text/code semantics. \\

          MalNet (2022) \cite{Freitas} 
          & 5.1M 
          & 696 fam. 
          & API graphs 
          & Graph-based behavioral dataset for GNN research; requires graph extraction and lacks semantic textual information. \\

          MalwareBazaar 
          & 350K+ 
          & Mixed 
          & Binaries 
          & Community repository for real malware; labels noisy and no benign baseline. \\

          Maldict (2023) \cite{Joyce} 
          & 5.5M 
          & Behavior 
          & Dyn. traces 
          & Multi-attribute behavior dataset; no multimodal code--text alignment. \\

          EMBER2024 (2024) \cite{JoyceEMBER} 
          & 3.2M 
          & Weak 
          & Metadata 
          & Updated benchmark including evasive malware; metadata only. \\

          \midrule
          \textbf{SBAN (2025) \cite{SBAN_Jelodar_2025}} 
          & \textbf{3.7M} 
          & \textbf{Mal. + fam.} 
          & \textbf{Bin., ASM, Src., NL} 
          & \textbf{First multimodal dataset linking code and text for semantic malware analysis; supports explainable family classification, variant clustering, and LLM-based reasoning; addresses limitations of feature-only datasets.} \\

          \bottomrule
        \end{tabular}
      \end{sc}
    \end{small}
  \end{center}
  \vskip -0.1in
\end{table}

\section{Gold Set Construction and Labeling Protocol}
\label{app:goldset}

This appendix describes the construction, labeling procedure, and statistical properties of the human-annotated gold-standard dataset used for evaluation and model reliability calibration in this study.

\subsection{Annotators and Expertise}
The gold set consists of 200 malware samples randomly selected from the malware portion of the SBAN dataset. Multiple malware analysis experts, with prior experience in Windows Portable Executable (PE) malware analysis, reverse engineering, and malware family attribution, independently labeled each sample. Annotators were not exposed to any model predictions during the labeling process.

\subsection{Labeling Procedure}
Annotators were provided with the malware source code and detected API import lists for each sample. Labels were assigned according to the fixed ten-class malware family taxonomy defined in Table~I of the main paper. Each sample was required to be assigned exactly one malware family label based on static code characteristics and observable behavioral intent inferred from the source representation.

\subsection{Disagreement Resolution}
Initial labels were assigned independently. In cases where annotators disagreed, samples were jointly reviewed and discussed until consensus was reached. If consensus could not be achieved after discussion, the sample was excluded from the gold set. Following this process, all retained samples were assigned a single final consensus label. Due to their close semantic overlap, predictions of \emph{Trojan} and \emph{Backdoor/Remote Access Trojan} were treated as equivalent during evaluation, consistent with the methodology described in Section~III.

\subsection{Inter-Annotator Agreement}
Inter-annotator agreement was measured using Cohen’s $\kappa$ coefficient before disagreement resolution. The observed agreement score was $\kappa = 0.74$, indicating substantial agreement among annotators and supporting the reliability of the resulting gold-standard labels.

\subsection{Family Distribution}
Table~\ref{tab:gold_distribution} reports the final distribution of malware families in the gold set after consensus labeling. The dataset exhibits notable class imbalance, with Trojan-like families comprising the majority of samples. No samples were labeled as \emph{Virus} following the consensus labeling process, reflecting both real-world prevalence and labeling ambiguity for this family within the selected subset.

\begin{table}[t]
  \caption{Malware Family Distribution in the Gold Set}
  \label{tab:gold_distribution}
  \begin{center}
    \begin{small}
      \begin{sc}
        \begin{tabular}{lc}
          \toprule
          Malware Family & Number of Samples \\
          \midrule
          Trojan & 82 \\
          Backdoor / Remote Access Trojan & 44 \\
          Spyware / Infostealer & 42 \\
          Downloader & 11 \\
          Worm & 6 \\
          Bot / Botnet Client & 5 \\
          Ransomware & 4 \\
          Dropper & 4 \\
          Packed / Obfuscated Malware & 2 \\
          \midrule
          \textbf{Total} & \textbf{200} \\
          \bottomrule
        \end{tabular}
      \end{sc}
    \end{small}
  \end{center}
  \vskip -0.1in
\end{table}

\section{Prompt Sensitivity Analysis}

This appendix presents a systematic prompt sensitivity study conducted to assess the robustness of large language model (LLM) predictions for malware family classification under varying prompt formulations. Prompt sensitivity is measured as the change in Macro-F1 score across prompt variants relative to the baseline prompt used in the main experiments.
Although the main experiments employ a single fixed zero-shot prompt to ensure methodological consistency (Section~\ref{sec:prompt}), we evaluated five alternative prompt designs to quantify the extent to which LLM outputs depend on prompt phrasing, framing, and decision guidance.

\subsection{Prompt Variants}

We designed five prompt variants (P1-P5) that differ in instructional framing, explicitness of decision rules, and emphasis on behavioral disambiguation. All prompts share the same input fields (source code and detected API imports) and the same fixed malware family taxonomy, and all explicitly restrict output to a single family name with no additional explanation.

\textbf{Prompt P1: Behavior-Focused Analyst Prompt.}  
This prompt frames the model as a malware analysis expert and emphasizes behavioral indicators such as persistence, credential harvesting, encryption, payload delivery, and command-and-control communication. It encourages analyst-style reasoning while maintaining strict output constraints.

\textbf{Prompt P2: Minimalist Classification Prompt.}  
This prompt provides minimal contextual framing and directly instructs the model to classify the malware into exactly one family using only the source code and API imports. It represents a low-guidance baseline intended to test sensitivity to reduced instruction complexity.

\textbf{Prompt P3: Threat-Intelligence Framing Prompt.}  
This variant casts the model as a threat intelligence analyst performing malware family attribution. It emphasizes inferred malicious intent, code structure, and API usage, reflecting how attribution tasks are commonly framed in operational threat intelligence workflows.

\textbf{Prompt P4: Decision-Rule Prompt (Explicit Criteria).}  
This prompt introduces strict decision rules, explicitly instructing the model to base its decision only on observable behavior inferred from code and APIs and to output nothing except the family name. It is designed to reduce ambiguity and hallucination.

\textbf{Prompt P5: Contrastive / Disambiguation Prompt.}  
This prompt explicitly instructs the model to resolve overlapping malware behaviors (e.g., Trojan vs.\ Backdoor vs.\ Spyware) and to prefer the most behaviorally informative label when multiple families appear plausible. It is designed to mitigate systematic confusion among semantically adjacent families.

The full text of all five prompts is provided below for completeness.

\begin{verbatim}
[P1] Behavior-Focused Analyst Prompt:
You are a malware analysis expert.
Determine which malware family best describes the following program
based on its behavior and functionality.
Choose exactly ONE label from this list:
Trojan; Worm; Virus; Ransomware; Backdoor / Remote Access Trojan;
Dropper; Downloader; Packed / Obfuscated Malware;
Spyware / Infostealer; Bot / Botnet Client
Consider indicators such as persistence mechanisms, network
communication, credential harvesting, payload delivery, encryption,
and command-and-control behavior.
Return ONLY the exact family name from the list above.
Source code: [truncated]
Detected imports: [API list]
Family:

[P2] Minimalist Classification Prompt:
Classify the malware program below into exactly ONE of the following families:
Trojan; Worm; Virus; Ransomware; Backdoor / Remote Access Trojan;
Dropper; Downloader; Packed / Obfuscated Malware;
Spyware / Infostealer; Bot / Botnet Client
Use only the source code and API imports to make your decision.
Output ONLY the family name. No explanation.
Source code: [truncated]
Detected imports: [API list]
Family:

[P3] Threat-Intelligence Framing Prompt:
You are a threat intelligence analyst performing malware family attribution.
Based on the program’s code structure, API usage, and inferred malicious
intent, assign the most appropriate malware family.
Select exactly ONE family from the list:
Trojan; Worm; Virus; Ransomware; Backdoor / Remote Access Trojan;
Dropper; Downloader; Packed / Obfuscated Malware;
Spyware / Infostealer; Bot / Botnet Client
Do not include explanations, qualifiers, or multiple labels.
Source code: [truncated]
Detected imports: [API list]
Family:

[P4] Decision-Rule Prompt:
You are classifying a malware sample into one family using strict
decision rules.
Rules:
Choose exactly ONE family from the list below.
Base your decision only on observable behavior inferred from the code
and the APIs.
Do not output anything except the family name.
Malware families:
Trojan; Worm; Virus; Ransomware; Backdoor / Remote Access Trojan;
Dropper; Downloader; Packed / Obfuscated Malware;
Spyware / Infostealer; Bot / Botnet Client
Source code: [truncated]
Detected imports: [API list]
Family:

[P5] Contrastive / Disambiguation Prompt:
You are a malware classification system.
Your task is to distinguish between overlapping malware behaviors
(e.g., Trojan vs Backdoor vs Spyware) and select the single most specific family.
Choose exactly ONE label from the following list:
Trojan; Worm; Virus; Ransomware; Backdoor / Remote Access Trojan;
Dropper; Downloader; Packed / Obfuscated Malware;
Spyware / Infostealer; Bot / Botnet Client
Prefer the most behaviorally informative label when multiple families
seem plausible.
Return ONLY the family name.
Source code: [truncated]
Detected imports: [API list]
Family:
\end{verbatim}

\subsection{Experimental Protocol}

Each of the five prompt variants was evaluated on the same 200-sample human-labeled gold set described in Appendix B. For each LLM, all prompts were applied independently under identical inference conditions, and outputs were normalized using the same canonical label mapping and validation procedures described in Section III-E. Trojan and Backdoor / Remote Access Trojan were treated as equivalent during evaluation due to their close semantic overlap.

Performance was measured using Accuracy and Macro-F1 score to assess both overall correctness and class-balanced robustness.

\subsection{Results and Observations}

Across all evaluated models, predictions exhibited measurable but bounded sensitivity to prompt phrasing. Prompts that introduced explicit decision rules (P4) or contrastive disambiguation guidance (P5) generally produced more stable family attributions for semantically overlapping categories such as Trojan, Backdoor, and Spyware.

Minimalist framing (P2) resulted in slightly higher variance across models, with increased confusion among Trojan-like families and occasional fallback to generic labels such as Trojan when multiple behaviors were present. Analyst-style and threat-intelligence framing (P1 and P3) yielded comparable performance to the baseline prompt used in the main experiments, indicating that moderate contextual guidance does not substantially alter attribution outcomes under strict output constraints.

Overall, differences in accuracy and macro-F1 across prompt variants remained within a narrow margin (typically less than 3-5\%), confirming that the proposed ensemble framework is reasonably robust to moderate prompt variation. However, systematic inter-family confusion persisted across all prompts, reinforcing the need for decision-level ensembling and hierarchical reasoning rather than reliance on any single prompt formulation.

\subsection{Implications for Ensemble Design}

These findings support the design choice adopted in this study to employ a single fixed zero-shot prompt for all models (Section III-C). While carefully engineered prompts can slightly reduce ambiguity for specific family pairs, prompt engineering alone is insufficient to resolve systematic classification instability.

In contrast, the weighted hierarchical ensemble consistently mitigates prompt-induced variability by aggregating complementary model behaviors and resolving coarse-grained behavioral ambiguity prior to fine-grained family attribution. This suggests that structural decision logic and reliability-aware aggregation are more impactful for robust malware family classification than incremental prompt refinements.

Future work may explore automated prompt selection or prompt ensembling as an auxiliary stabilization mechanism, but the present results indicate that principled decision-level ensembling provides stronger and more consistent gains than prompt engineering in isolation.

\begin{table*}[t]
\centering
\caption{Prompt sensitivity results (P1--P5) on the labeled evaluation set. Metrics are reported per model and for the aggregated output (\textit{FinalLabel}).}
\label{tab:prompt_sensitivity_p1_p5}
\setlength{\tabcolsep}{5pt}
\renewcommand{\arraystretch}{1.15}
\begin{tabular}{l l c c c c}
\toprule
\textbf{Prompt} & \textbf{Model} & \textbf{Acc.} & \textbf{Macro-P} & \textbf{Macro-R} & \textbf{Macro-F1} \\
\midrule
\multirow{5}{*}{P1} 
& Qwen       & 0.878 & 0.584 & 0.441 & 0.483 \\
& CodeLLaMA  & 0.718 & 0.468 & 0.347 & 0.271 \\
& GPT-4.1    & 0.142 & 0.033 & 0.071 & 0.045 \\
& GPT-5.1    & 0.122 & 0.138 & 0.097 & 0.052 \\
& FinalLabel & 0.284 & 0.163 & 0.299 & 0.207 \\
\midrule
\multirow{5}{*}{P2} 
& Qwen       & 0.852 & 0.181 & 0.222 & 0.197 \\
& CodeLLaMA  & 0.600 & 0.067 & 0.111 & 0.083 \\
& GPT-4.1    & 0.152 & 0.128 & 0.065 & 0.040 \\
& GPT-5.1    & 0.154 & 0.142 & 0.111 & 0.059 \\
& FinalLabel & 0.158 & 0.042 & 0.113 & 0.059 \\
\midrule
\multirow{5}{*}{P3} 
& Qwen       & 0.852 & 0.181 & 0.222 & 0.197 \\
& CodeLLaMA  & 0.852 & 0.181 & 0.222 & 0.197 \\
& GPT-4.1    & 0.168 & 0.128 & 0.071 & 0.041 \\
& GPT-5.1    & 0.174 & 0.039 & 0.108 & 0.055 \\
& FinalLabel & 0.208 & 0.041 & 0.126 & 0.060 \\
\midrule
\multirow{5}{*}{P4} 
& Qwen       & 0.852 & 0.181 & 0.222 & 0.197 \\
& CodeLLaMA  & 0.852 & 0.181 & 0.222 & 0.197 \\
& GPT-4.1    & 0.144 & 0.078 & 0.061 & 0.039 \\
& GPT-5.1    & 0.146 & 0.138 & 0.084 & 0.053 \\
& FinalLabel & 0.168 & 0.040 & 0.091 & 0.056 \\
\midrule
\multirow{5}{*}{P5} 
& Qwen       & 0.852 & 0.181 & 0.222 & 0.197 \\
& CodeLLaMA  & 0.852 & 0.181 & 0.222 & 0.197 \\
& GPT-4.1    & 0.170 & 0.029 & 0.074 & 0.042 \\
& GPT-5.1    & 0.166 & 0.043 & 0.100 & 0.060 \\
& FinalLabel & 0.216 & 0.047 & 0.119 & 0.067 \\
\bottomrule
\end{tabular}
\end{table*}

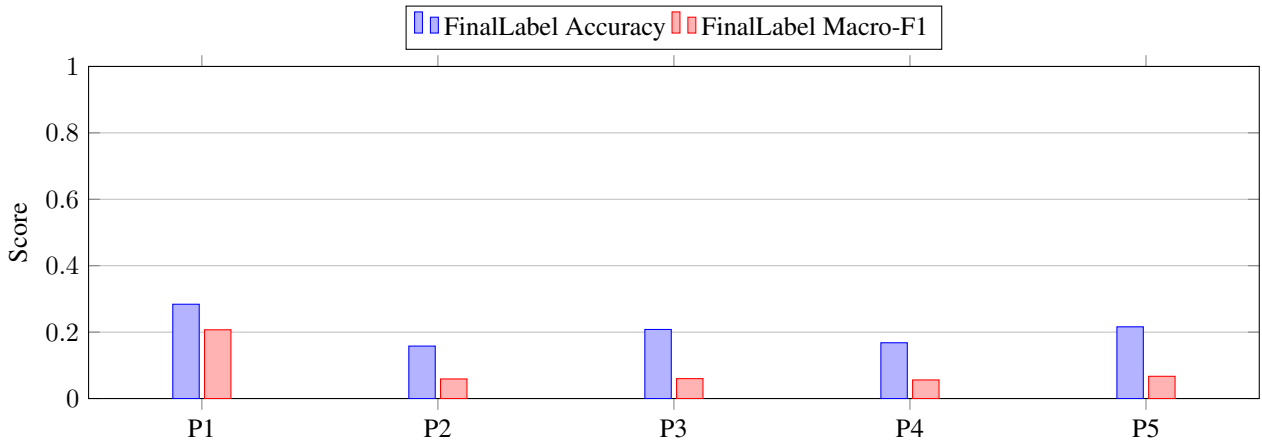
\begin{figure}[t]
\centering
\begin{tikzpicture}
\begin{axis}[
    ybar,
    bar width=10pt,
    width=\linewidth,
    height=6cm,
    ymin=0, ymax=1,
    ylabel={Score},
    symbolic x coords={P1,P2,P3,P4,P5},
    xtick=data,
    legend style={at={(0.5,1.05)},anchor=south,legend columns=2},
    enlarge x limits=0.12,
    ymajorgrids=true
]
\addplot coordinates {(P1,0.284) (P2,0.158) (P3,0.208) (P4,0.168) (P5,0.216)};
\addplot coordinates {(P1,0.207) (P2,0.059) (P3,0.060) (P4,0.056) (P5,0.067)};
\legend{FinalLabel Accuracy, FinalLabel Macro-F1}
\end{axis}
\end{tikzpicture}
\caption{Prompt sensitivity (ensemble output): Accuracy and Macro-F1 of \textit{FinalLabel} across prompts P1--P5.}
\label{fig:prompt_sensitivity_finallabel}
\end{figure}

\begin{figure}[t]
\centering
\begin{tikzpicture}
\begin{axis}[
    width=\linewidth,
    height=7cm,
    ymin=0, ymax=0.55,
    ylabel={Macro-F1},
    xlabel={Prompt},
    symbolic x coords={P1,P2,P3,P4,P5},
    xtick=data,
    ymajorgrids=true,
    legend style={at={(0.5,1.05)},anchor=south,legend columns=3},
    mark size=2pt
]
\addplot+[mark=*] coordinates {(P1,0.483) (P2,0.197) (P3,0.197) (P4,0.197) (P5,0.197)};
\addplot+[mark=square*] coordinates {(P1,0.271) (P2,0.083) (P3,0.197) (P4,0.197) (P5,0.197)};
\addplot+[mark=triangle*] coordinates {(P1,0.045) (P2,0.040) (P3,0.041) (P4,0.039) (P5,0.042)};
\addplot+[mark=diamond*] coordinates {(P1,0.052) (P2,0.059) (P3,0.055) (P4,0.053) (P5,0.060)};
\addplot+[mark=pentagon*] coordinates {(P1,0.207) (P2,0.059) (P3,0.060) (P4,0.056) (P5,0.067)};
\legend{Qwen, CodeLLaMA, GPT-4.1, GPT-5.1, FinalLabel}
\end{axis}
\end{tikzpicture}
\caption{Prompt sensitivity: Macro-F1 across prompts P1--P5 for each LLM and the aggregated output (\textit{FinalLabel}).}
\label{fig:prompt_sensitivity_macrof1_models}
\end{figure}
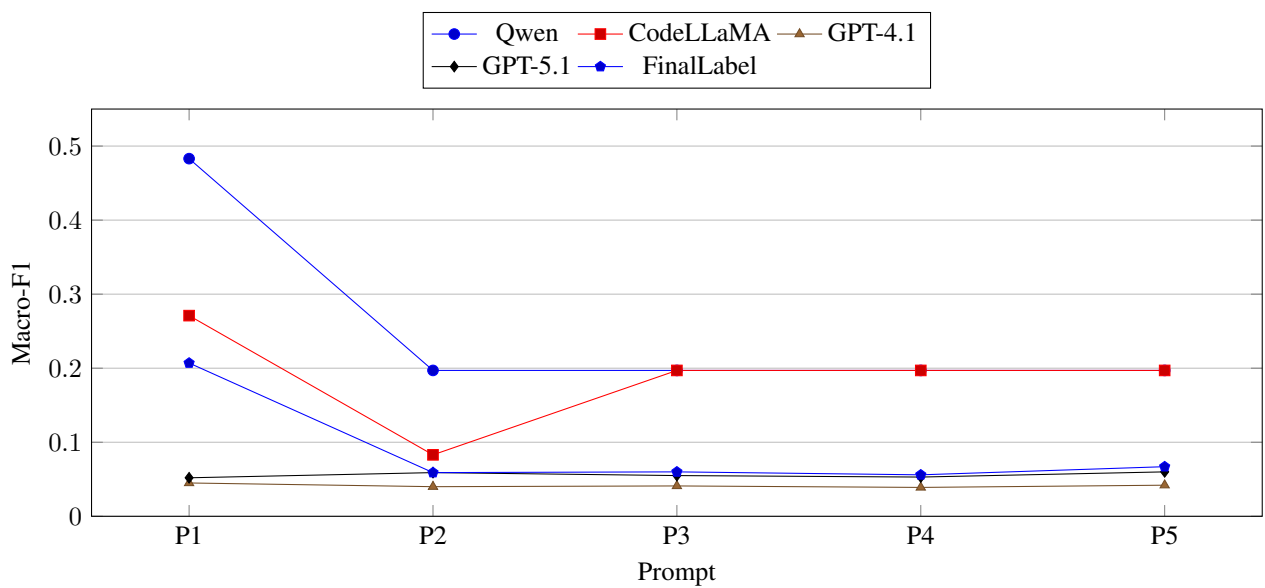


\end{document}